# SINGLE PAGE APPLICATION AND CANVAS DRAWING

Renien John Joseph

Zone24x7 (Private) Limited460, Nawala Road,Koswatte, Sri Lanka

*ABSTRACT*

*Recently, with the impact of AJAX a new way of web development techniques have been emerged. Hence, with the help of this model, single-page web application was introduced which can be updated/replaced independently. Today we have a new challenge of building a powerful single-page application using the currently emerged technologies. Gaining an understanding of navigational model and user interface structure of the source application is the first step to successfully build a single- page application.*

*In this paper, it explores not only building powerful single-page application but also Two Dimensional (2D) drawings on images and videos. Moreover, in this research it clearly express the findings on 2D multi-points polygon drawing concepts on client side; real-time data binding in between drawing module on image , video and view pages.*

*KEYWORDS*

*AJAX, Two Dimensional, Multi-point Polygon, Data binding*

## 1. INTRODUCTION

The World Wide Web (WWW) is constantly evolving and so is the way developers write the applications that runs upon it. Web applications are popular due to the ubiquity of web browsers, and it convinces the entire user to solve their respective domain problem in any time. Many factors need to consider when developing web application [10]. Most of the developers should always consider many factors before undertaking a web project. Neilsen (2005) [10], have clearly mentioned about some valuable points that need to be considered to develop a web application [18][1].

- Target user or the audience
- Main purpose of the site
- Understand the domain of the site
- Scalability
- Planning [5]
- Search Engine Optimization

Interaction in classic web applications is based on a multi-page interface model, in which for every request the entire interface is refreshed.

Single Page Applications (SPA) is built on expanding reach via the browser, reducing round tripping, and enhancing User Experience (UX). SPA is also known as Single-Page Interface. SPA is able to be composed due to the new emerged technology, dubbed AJAX (Asynchronous JavaScript and XML) [6]. Khapre and Chandramohan (2011) [7] have clearly emphasized the importance of web services due to the immerged technology and to have a significant impact on different domain. SPA is composed of individual page that can be updated independently on each user's action, so that the entire page does not need to be reloaded like classical web application. This, in turn, helps to increase the levels of interactivity, responsiveness and user satisfaction [2].

Currently the web technology growth is in its peak level, to compose powerful SPA. It is all up to the list of web technologies including with client side technology [3]. But at the end clients rate web application according to client side performance. So, for a SPA the client side technology carries a great value. The success of real-time application depends on selection of appropriate platform as well as implementing the right information management strategy.

In this paper, it evaluates on client side framework to compose SPA along with it innovative solution to increase the user experience. It also includes manipulating server side data along with client side data and on design, implementation section it is been clearly draft on the findings and solutions.

Angular.js, an open source JavaScript framework have been developed that enable and gives an extreme freedom to client-side developers to build powerful SPA. This framework uses Model-View-Controller (MVC) architecture [8], data binding, client-side templates and dependency injection [13] to create a much-needed structure for building web apps. While reading further with the paper, it covers a mechanism to draw two dimensional (2D) polygons [9] on images and video frames.

## 2. RELATED WORK

The web has become far more different place in present when comparing in six year back. Twitter and Facebook are one of the most popular social media sites. As more and more people use these enterprise applications since it was built with different user experience that does not look old and slow when comparing with primary stage of development. This massive growth has been taken placed in mobile development as well since the web technology has been intersecting with desktop and mobile plat form too [19]. Etter (2013) [4] recently on his magazine has given is thoughts about the real-time web application development on mobile platform. As a traditional web development pattern mostly opens source technologies and ASP.NET MVC4 mobile customized version that can render the page effectively. Both of these techniques follow the traditional Web development pattern where each page is focused on a specific task [19].

But at currently due to the power of HTML5 and JavaScript the web development has been totally changed [12]. Etter (2013) [4] has highlighted main two benefits using SPA; less network bandwidth and faster navigation. Here we can be sure the first point is accomplished because, though the initial page is larger the users are essentially getting multiple pages in a single page and this will be transmit supporting navigation by JavaScript libraries.

So, on-page navigation may need little or no request between the client and the server. Faster navigation is the second benefit. Any required data (typically JSON-formatted) may be retrieved asynchronously using JavaScript.

International Journal of Web & Semantic Technology (IJWesT) Vol.6, No.1, January 2015

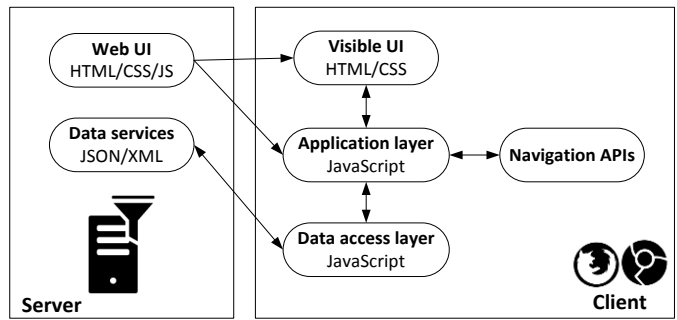

Figure 1 High level architecture of SPA

The Figure 1 show a clear picture of Typical Web architecture contains a server and client, where server contains an end point to server HTML/CSS and JS. The client side is being rendered as Visible UI and contains some JavaScript as well that is web technology.

The canvas rendering module which author mentioned in introduction section deals with client side rendering. It includes 2D multi-point polygon shape editor on image and video canvas. The module has so many different features but the most unique one is to bound the multi-point polygon inside the drawing canvas during real-time HTML DOM rendering.

Continuously it has been a challenging task to researchers to deals with multi point shapes. Many researchers have worked on 3D and 2D multi-point shape rendering for game development [14]. In computer graphics algorithm, real-time Bounding Box area is one of the popular research points for very long time.

Since the technology was moving forward to cloud the researchers were started to work on adapting these algorithms and solutions to client side development. It was so challenging because the performance factor really means on client side development. Tammik (2013) [16], on his article have argued on 3D collision box where by elimination the Z-axis can easily adopt the algorithm to 2D collision box.

In 2D multi point polygons (Figure 2), the following expression p=(x+y) and q=(x-y) correspond to lines with slopes of (–1) and 1. In an article [15] have clearly pointed out the different geometrical equation for 3D axis and 2D axis. For a 2D polygon points set, using the Equation 1 can compute the minimum and maximum over the collection of points from the 2D polygon shape model.

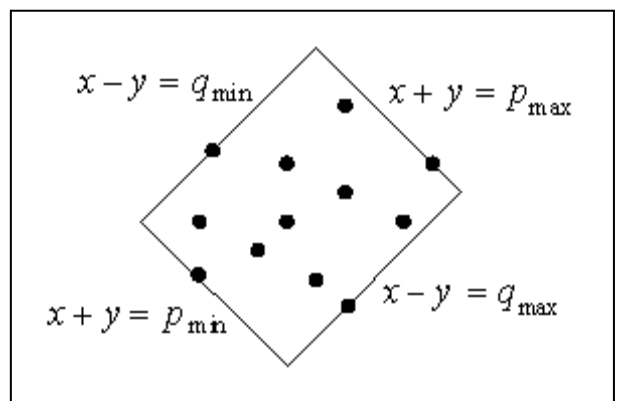

**Figure 2 Polygon points**

$$p=(x+y) \text{ and } q=(x-y) \quad \text{- Equation 1}$$

## 3. WORK FLOW OF THE PROTOTYPE

To make sure and to evaluate the current technology the product was developed with lot more features. It has mainly three modules. Figure 3 clearly shows the high level details of the modules. Angular.js framework is used to develop the data binding and it helps to develop a power full web application at the end.

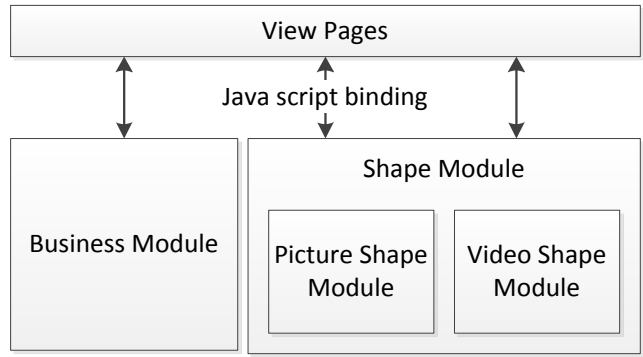

Figure 3 over all modules of the system

### 3.1 Business Module

The business module will differ depending on the domain. Along with the server side business modules most of the client side module design and the pattern can have the same structure, it is not a must but it depends on the view pages.

### 3.2 Shape Module

This module was developed using a JavaScript framework called KineticJS. KineticJS is a JavaScript library that helps to simplify the development of interactive HTML5 applications very effectively. In this case the built interactive application is different when comparing with 2D drawing applications such as games, image editors, interactive maps and cartoons. From a technical point of view it is an extension to the HTML5 and Canvas 2D context in the form of a JavaScript library [17].

The shape module breaks in to several other modules such as rectangles and polygons. These modules are developed in such way where it can be used as a plugin for a web application. Secondly it is developed as extendible framework where the developers can do the improvement and add more valuable features without affecting the existing shape module.

### 3.3 View Pages

Angular.js has been designed in a way where each view page is directly binding to a JavaScript controller file. The controllers hold the logic; to retrieve the model, to trigger different operations to perform on it and to transform the model to show the details on the view page. Along with that responsibility of validation, making server calls, bootstrapping the view with the right data, and mostly everything in between model and view belongs on controller (Figure 4). In addition, this module will work on all browsers that support HTML5, and even on mobile devices.

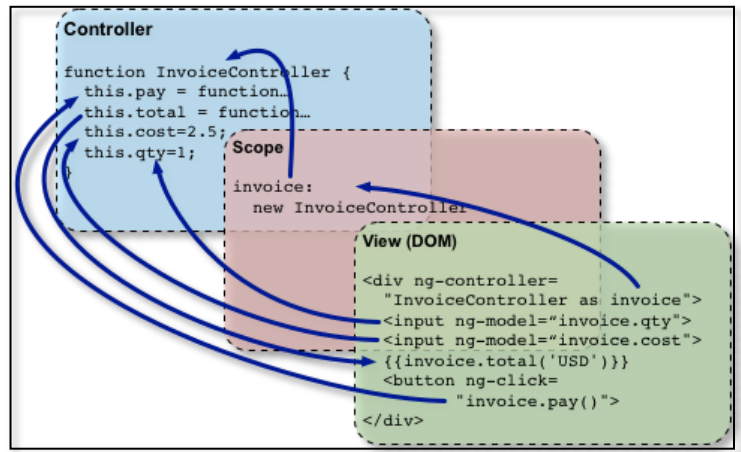

Figure 4 Angular.js Binding

## 4. IMPLEMENTATION
### 4.1 Shape Module

It is structured and implemented in a way to work on both image and video canvas. In the Figure 5 it clearly expresses, each main shape is consisting of different polygons. During the development process came across a problem to bound the 2 dimensional shape polygons inside the image and video. To solve this issue need to implement an algorithm (Equation 1) where it involves creating a virtual 2D bounding box surrounding the multi point's polygon shape ( Figure 6).

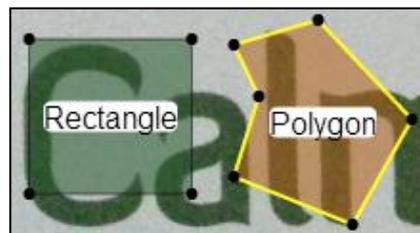

**Figure 5 Polygon Shapes**

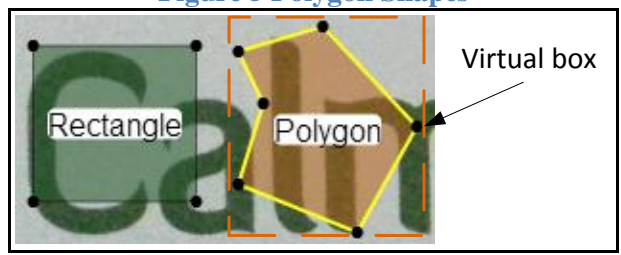

Figure 6 Virtual Bounding Box

It was so easy to deal the real time web interface using the Angular.js where it always binds with the JSON value. As previously mentioned shape module is developed with the help of Kinectic.js where in allowed the freedom to seamlessly integrate on the view page model with the help of Angular.js.

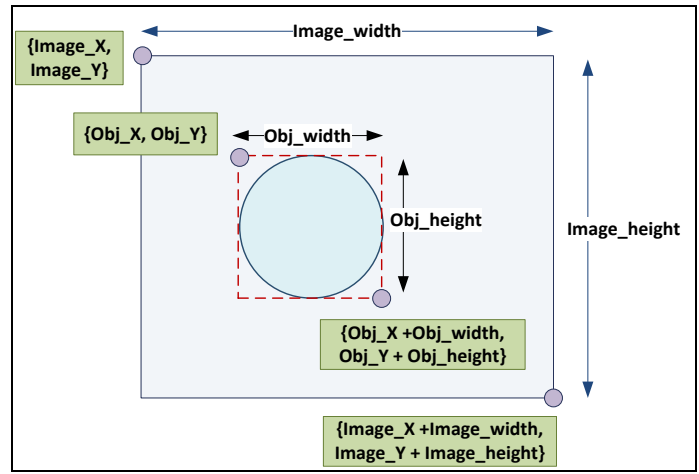

Figure 7 Bound inside image

The following
Figure 8 shows how easy it was to bind the data in between different layers. After calculating the virtual box according to
Figure 7 the mathematical logic algorithm was implemented which helped to bound any multi-points polygon inside images and videos successfully.

Since author was working on two different libraries of layers it could lead to break the shape module. But Angular.js supports a great feature where easily can bind any library data very effectively. Because of this integration the users had great experience where it allowed to refresh and produce the result without changing the pages (
Figure 6,
Figure 9).

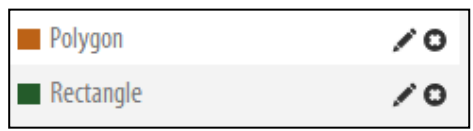

Figure 8 Binding in between two layers

Figure 9 Binding between drawing canvas and view page

## 4.2 View Module

Directives in AngularJS are used to make custom HTML elements and simplify DOM manipulation. They can modify the behavior of new and existing DOM elements, by adding custom functionality, like a date picker or an auto complete widget. Using this great feature it allows and gives freedom to add more custom HTML DOM elements that can solve the problem very effectively which gives a really good user experience. The framework has been written in such a way that developer can try out with their own innovative ideas. Many problems have been able solved on the view model and one of the most important finding was binding the CSS class. Basically to manipulate with the directives initially create a separate customized templates. Since it allowed not only data binding but and also CSS binding, was able to dynamically change the CSS class and render the several components on the same template in real time on the same page. Hence, this allowed to reduce the development time and also was able to handle the no of files on the project so efficiently. More over the end result was so spectacular; it really increases the user experience level.

## 5. TESTING

All core modules were tested with different relevant input to ensure all the proposed functionalities were working. To ensure a flawless end product we carried out two types of testing as follows.

1. Shape Module (Performance)
2. Business Module (User Experience)

This process was tested under the following hardware and OS requirements.

- Laptop (Intel(R) Pentium(R) Dual CPU T3400 @ 2.16GHz (2 CPUs))
- Window 8
- Chrome 16

### 5.1 Shape Module

The testing results proved while the number of polygon shape input was increased the accuracy or robust level of rendering the 2D canvas stays the same but the rendering time increases with a slight linear growth. But to increase the computation speed it is also depend on the browser and the computer. The
Figure 10 in the graphs show the clear picture that even with the increasing no of the data it is not badly affected with the rendering time.

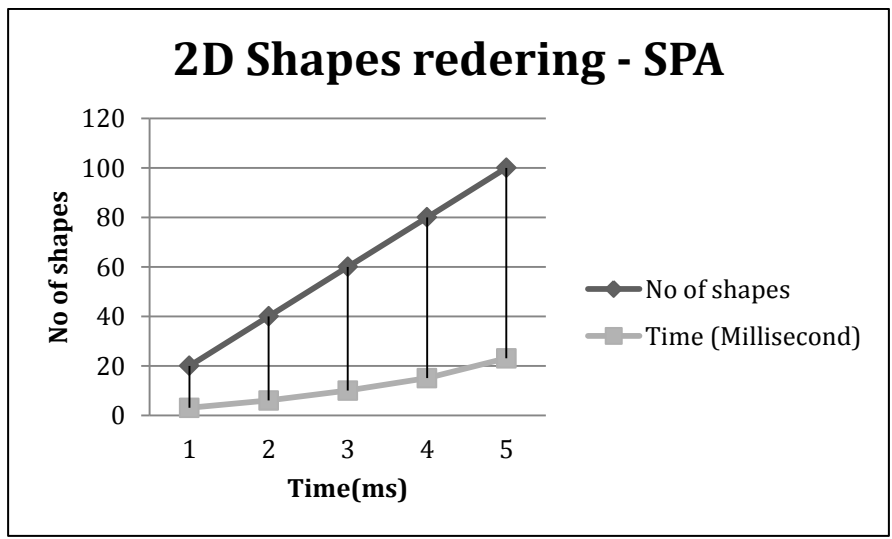

Figure 10 Shape module test result

### 5.2 User Experience

For SPA most important thing is end user experience. It is an acceptable truth comparing to classical web application, single page application is so high rated to user experience level. This web application was tested with different age group people and they were so existed with the end result. The (
Figure 11,
Figure 12) following graph clearly shows, that the SPA is great success where it doesn't shows a big variant on the user experience level. The testing was carried out with a bunch of people selected people with technical skilled people and also different age period people.

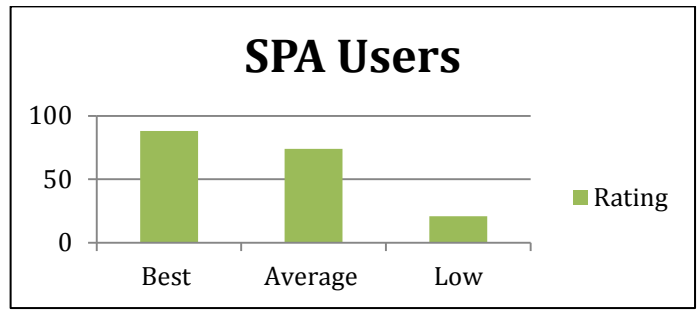

Figure 11 SPA users rating

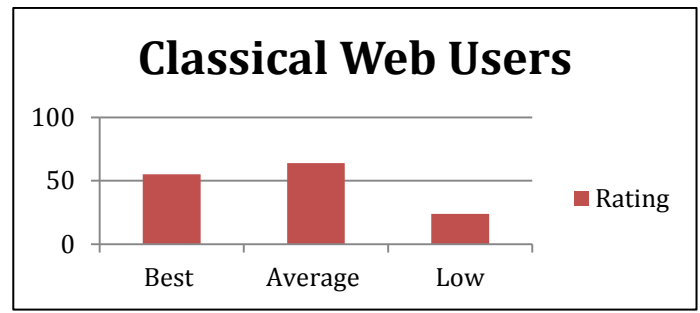

Figure 12 Classic web users rating

## 6. CONCLUSION AND RECOMMENDATIONS

The main contribution of this paper is the correct path to building a single page application, drawing on images and videos very effectively.

Since the current web technology growth is very high; it gives a freedom to render very complex computations on client side browsers like image processing, 2D and 3D rendering very effectively [11]. Due to the seamless integration or biding in between the technologies the rendering of the page is very smooth. The user experiences of the pages are so high because during the testing it was clear that it renders the pages along with the data on the client side. It is really promising that clients would not get disappointment when using the single page application.

## ACKNOWLEDGEMENTS

I express my deepest and most sincere gratitude to all of them who helped me in the stage of requirement gathering and evaluation, without all their support this project would not have been successful.

**Authors**
**Renien Joseph**

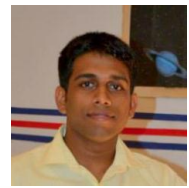

Renien Joseph is a Software Engineer currently working in Zone 24X7 Pvt (Ltd). An active developer with a very broad knowledge and keen interests towards Big Data, 3D visualization, Computer vision technologies, Microsoft technologies, modern web technologies and mobile technologies.